\documentclass[aps,12pts,twocolumn,nofootinbib,showpacs,superscriptaddress]{revtex4-1}

\usepackage{graphicx}
\usepackage{dcolumn}
\usepackage{tabu}
\usepackage{amsmath,amssymb}
\usepackage{float}
\usepackage{natbib}
\usepackage{balance}
\usepackage{mathrsfs}
\usepackage{bm}
\usepackage{lipsum}
\usepackage[dvipsnames]{xcolor}
\usepackage[title]{appendix}
\usepackage[breaklinks=true,pdfborder={0 0 0},bookmarksopen]{hyperref}
\hypersetup{
	colorlinks   = true, 
	urlcolor     = blue, 
	linkcolor    = blue, 
	citecolor   = red 
}

\begin{document}

\title{Black hole mechanics and thermodynamics in the light of Weyl transformations}

\author{Fay\c{c}al Hammad} \email{fhammad@ubishops.ca} 
\affiliation{Department of Physics and Astronomy \& STAR Research Cluster, Bishop's University, 2600 College Street, Sherbrooke, QC, J1M~1Z7 Canada} 
\affiliation{Physics Department, Champlain 
College-Lennoxville, 2580 College Street, Sherbrooke,  
QC, J1M~0C8 Canada}

\author{\'Etienne Mass\'e}
\email[]{Etienne.Masse@usherbrooke.ca}
\affiliation{Department of Physics and Astronomy \& STAR Research Cluster, Bishop's University, 2600 College Street, Sherbrooke, QC, J1M~1Z7
Canada} 
\affiliation{D\'epartement de Physique, Universit\'e de Sherbrooke, Sherbrooke, QC, J1K~2X9 Canada}

\author{Patrick Labelle}
\email[]{plabelle@crc-lennox.qc.ca}
\affiliation{Physics Department, Champlain 
College-Lennoxville, 2580 College Street, Sherbrooke,  
QC, J1M~0C8 Canada}

\begin{abstract}
The fate of black hole thermodynamics under Weyl transformations is investigated by going back to the laws of black hole mechanics. It is shown that the transformed surface gravity, that one would identify with the black hole temperature in the conformal frame, as well as the black hole entropy, that one would identify with the horizon area, cannot be invariant. It is also shown that the conformally invariant surface gravity, attributed to the so-called ``conformal Killing horizon", cannot represent the black hole temperature in the conformal frame. Finally, using familiar thought experiments, we find that the effect a Weyl transformation should have on black hole thermodynamics becomes even subtler than what is suggested by the laws of black hole mechanics.   
\end{abstract}

\pacs {04.70.Bw, 04.70.Dy, 04.20.-q}
\maketitle
\section{Introduction}\label{SecI}
Black hole mechanics, discovered in the seminal paper \cite{BCH}, suggests a complete analogy between the physics of spacetime and the laws of thermodynamics. Although such a discovery was based on the {\it classical} notions of spacetime, as dictated by general relativity, the subsequent landmark paper of Hawking \cite{Hawking} turned this analogy into a true identity based on the effect of spacetime curvature on {\it quantum} matter fields. Hawking temperature, in turn, implies that black hole horizons are reservoirs of tremendous entropy. This remarkable fact was first suggested in Bekenstein's pioneering papers \cite{Bekenstein1,Bekenstein2} using thought experiments that involved {\it both} matter and spacetime. It is therefore of great importance to investigate deeper this surprising connection between the two.

Now, the meaning at the classical level of such results has already been unraveled by various authors, by showing that the origin of spacetime thermodynamics might be traced back to diffeomorphism symmetry in gravitational physics \cite{WaldEntropy,IyerWald,JacobsonKangMyers}. However, we believe that one might still achieve a deeper understanding of the intimate connection between the classical spacetime and the quantum mechanical matter that leads to black hole thermodynamics by a {\it non-separate} investigation of both entities.

Gravitational physics is about the interaction between spacetime and matter. Curiously though, this interaction, as given by Einstein equations, still displays a fundamental dichotomy between matter and spacetime. The latter is the purely geometric quantity on the left-hand side of the field equations, while the former is the non-geometric quantity on the right-hand side of the equations. It is therefore remarkable that concepts like temperature and entropy, whose origins are deeply rooted in our concepts of matter and its behavior, suddenly become identified with (or at least assigned to) geometry. Given this fundamental dichotomy between matter and spacetime, however, one way to learn about their mutual interaction is actually to use Weyl conformal transformations or ``mappings". These are transformations that take the original metric $g_{ab}$ of spacetime to a new metric $\tilde{g}_{ab}$ given by, 
\begin{equation}\label{Weyl}
\tilde{g}_{ab}=\Omega^2g_{ab}.
\end{equation}
The factor $\Omega(x)$ being an arbitrary spacetime-dependent everywhere regular, non-vanishing, and smooth function. Such a factor is usually chosen to be equal to unity at infinity in order to preserve the asymptotic structure of spacetime, and hence of the distant observer, away from the central object under study.

In fact, a Weyl transformation affects both spacetime {\it and} the matter it contains, but in {\it different} ways. This fact is precisely the origin of the well-known conformal non-invariance of Einstein equations. As such, we believe that Weyl conformal mapping constitutes a very useful tool in investigating any concept of spacetime physics, including black hole thermodynamics. Indeed, this observation has actually been used in Refs.~\cite{Hammad1,Hammad2} to understand the true nature of some of the well-known concepts of quasilocal energies found in the literature. Thanks to Weyl's transformation, it was found there that the well-known Misner-Sharp and Hawking-Hayward quasilocal masses represent in fact the {\it geometric equivalent} of a mass. As such, any attempt to identify these with real ``material'' masses would automatically fail in any investigation that needs to bring in spacetime and matter on an equal footing. In other words, Weyl conformal transformations have been used in those references, not as a mere exercise for checking the conformal (non-)invariance of a physical quantity like quasilocal mass, but instead as a tool to probe into our deepest views about spacetime and matter. It just turns out that there is probably no better place where spacetime and matter are intertwined deeper than in the field of black hole thermodynamics. Thus, our use of conformal transformations in the present paper is done in the same spirit as in Refs.~\cite{Hammad1,Hammad2}: Not as a mere exercise, but as a tool to shed more light on the mysteries of black hole thermodynamics. 

Various authors have investigated the conformal behavior of Hawking temperature and Bekenstein-Hawking entropy, either ({\it i\,}) based on the classical definitions of surface gravity and black hole horizons \cite{JacobsonKang,FaraoniPRD, FaraoniGalaxies,Majhi,NielsenShoom,NielsenFirouzjaee}, or ({\it ii\,}) \cite{JacobsonKang,FaraoniPRD,Majhi,NielsenFirouzjaee,BhattacharyaMajhi,BhattacharyaDasMajhi} based on the various methods devised in the literature for recovering those famous formulas \cite{ParikhWilczek,CarlipCQG,RobinsonWilczek,MajhiPadmanabhan} (see also the nice review in Ref.~\cite{CarlipReview}). Various authors came to the conclusion that Hawking temperature is conformally invariant. It is likely, however, that any conclusion based on  ``sophisticated" methods, devised to work in the original frame (also known as the Jordan frame), might very well be incomplete. Indeed, as investigated recently in Refs.~\cite{FaraoniPrainZambrano,Hammad3}, under a Weyl transformation, a black hole might disappear altogether in the conformal frame (also known as the Einstein frame). In the absence of any horizon, one wonders how one could still talk about temperature, let alone an invariant temperature. As we shall argue, however, the issue is not in adopting a ``sophisticated'' method vs. adopting a ``simple'' method. Rather the issue lies at a more fundamental level which consists of the difference between matter and geometry. Therefore, instead of assessing these methods, or even appealing to them as done by previous authors interested in the conformal behavior of black hole thermodynamics, we chose here to tackle the problem from a different angle. This will allow us to see the issue from a novel perspective and a general point of view.

In this regard, Ref.~\cite{JacobsonKang}, which was among the first to suggest the conformal invariance of the black hole temperature, is of particular interest to us here. In fact, the authors in that reference argued for the conformal invariance of Hawking temperature by first using the classical concept of surface gravity. The authors then corroborated their result by applying the ``sophisticated" method of trace anomaly in the conformal frame. However, showing that surface gravity is conformally invariant there came at the price of introducing a new notion of horizon; the so-called ``conformal Killing horizon" \cite{DyerHonig,SultanaDyer}. 

Our aim in this paper is to investigate the conformal behavior of black hole thermodynamics by going back to the original laws of black hole mechanics instead, and checking the status of these laws in the conformal frame. As far as we know, going back to the laws of black hole mechanics in order to investigate the conformal issue of black hole thermodynamics has not been done previously in the literature. From this investigation, it easily follows, as we will see, that the surface gravity cannot be conformally invariant. Therefore, if one is willing to keep in the conformal frame this identification of the surface gravity with the black hole temperature, then the latter cannot be conformally invariant either. Furthermore, we find that the surface gravity associated with the so-called ``conformal Killing horizon" cannot be uniform over the horizon in the conformal frame, making the identification of such a surface gravity with temperature problematic.

Moreover, we show that the above-mentioned fundamental dichotomy between spacetime and matter is best revealed through the familiar thought experiments, built around a black hole, when performed in the conformal frame. Such thought experiments show, indeed, that black hole thermodynamics curiously becomes conformally invariant and, hence, reveal that the conformal behavior of black hole thermodynamics is much subtler than what the conformal behavior of the laws of black hole mechanics alone implies. In fact, as we shall see, our results are so far from being trivial that one should not even expect to recover the familiar area law of Bekenstein-Hawking entropy in the conformal frame. As we shall see, however, these seemingly contradictory results are all due to the fundamental difference between spacetime and matter, which thus requires one to first specify what one is interested in finding the conformal behavior of. As such, our present work just does to black hole thermodynamics what has been done to quasilocal masses in Refs.~\cite{Hammad1,Hammad2}: It filters out purely geometric concepts from purely material ones.

The outline of the remainder of this paper is as follows. In Sec.~\ref{SecII}, we briefly recall the four laws of black hole mechanics and the main steps towards their derivations. In Sec.~\ref{SecIII}, we derive the condition for having a true Killing vector in the conformal frame and investigate the corresponding four laws of black hole mechanics. In Sec.~\ref{SecIV}, we show that the surface gravity, obtained from the so-called ``conformal Killing vector", is not uniform over the horizon and, hence, cannot be identified with the temperature of such a black hole horizon in the conformal frame. In Sec.~\ref{SecV}, we discuss thought experiments and argue that, under a Weyl transformation, both the horizon entropy and the black hole temperature are invariant but that, nevertheless, this does not contradict what is found based on black hole mechanics. We end this paper with a brief discussion and conclusion section.
\section{The four laws of black hole mechanics}\label{SecII}
In this section we briefly recall some of the main mathematical tools used in the physics of black holes, the four laws of black hole mechanics as first formulated in Ref.~\cite{BCH}, and outline their derivations that we are going to retrace in detail in Sec.~\ref{SecIII}. 

In this paper we are interested in stationary spacetimes. A stationary black hole spacetime possesses a Killing vector field $\xi^a$ such that the Lie derivative $\pounds_\xi g_{ab}$ of the metric $g_{ab}$, along the direction $\xi^a$, is identically zero\footnote{Recall that a stationary black hole is either static or axially symmetric \cite{Hawking1972}.}. To be a Killing vector field, such a field $\xi^a$ must satisfy, in terms of the covariant derivative, the following Killing equation:
\begin{equation}\label{KillingEq}
\nabla_{(a}\xi_{b)}=0.
\end{equation}
Here and henceforth, round (square) brackets on indices mean symmetrization (anti-symmetrization) on those indices with the usual factor of $\frac{1}{2}$. When a Killing vector field becomes null on the horizon, such a horizon is called a Killing horizon. The event horizon of a stationary black hole is a Killing horizon. 

Another important quantity is the surface gravity $\kappa$. It represents the force --- as measured by a distant observer --- that needs to be exerted on a unit mass to keep it stationary at the event horizon of a black hole\footnote{Throughout the paper, we are going to use the natural units, in which $G=\hbar=c=1$.}. It is defined by,
\begin{equation}\label{SurfaceGrav1}
\xi^a\nabla_a\xi^b=\kappa\xi^b.
\end{equation}
A very important note to make here is that, thanks to Killing's equation (\ref{KillingEq}), one can also extract from the definition (\ref{SurfaceGrav1}) the following equivalent expressions for the surface gravity on the Killing horizon of a stationary black hole (see e.g. Ref.~\cite{Wald}):
\begin{align}\label{SurfaceGrav2}
\nabla_a(\xi^b\xi_b)&=-2\kappa\xi_a,\\\label{SurfaceGrav3}
(\nabla^a\xi^b)(\nabla_a\xi_b)&=-2\kappa^2.
\end{align}
It is the surface gravity of the black hole horizon that is identified --- up to a numerical factor --- with the black hole's temperature. Since the temperature of any physical object in thermodynamic equilibrium should be uniform over the body's surface, one also requires such a feature from the surface gravity of a stationary black hole. This leads to the zeroth law.  

\subsection{Zeroth law in Jordan frame}
The zeroth law of black hole mechanics states that the surface gravity of a stationary black hole is uniform over the entire event horizon. The derivation of this result is lengthy but otherwise straightforward. The key result in such a derivation is to show that \cite{Wald},
\begin{equation}\label{0Law}
\xi_{[a}\nabla_{b]}\kappa=0.
\end{equation}
The key steps leading to such a result heavily rely on Killing's equation (\ref{KillingEq}), but, very importantly also, do not require the satisfaction of Einstein equations as shown in Ref.~\cite{RaczWald}. In fact, all one needs to show is that, on one hand, $\xi_{[a}\nabla_{b]}\kappa=-\xi_{[a}{R_{b]}}^c\xi_c$ and on the other, $-\epsilon_{abcd}\xi^{[c}{R^{d]}}_e\xi^e=\nabla_{[a}\omega_{b]}$, where $\omega_a$ is the twist one-form defined by $\omega_a=\epsilon_{abcd}\xi^b\nabla^c\xi^d$. This would imply that, $\xi_{[a}\nabla_{b]}\kappa=-\frac{1}{4}\epsilon_{abcd}\nabla^{[c}\omega^{d]}$. Then, by using the fact that for a Killing vector field we have $\omega_a=0$ all over the Killing horizon, identity (\ref{0Law}) would immediately follow. 
\subsection{First law in Jordan frame}\label{sec:IIB}
The first law of black hole mechanics relates the variation of the black hole's mass to the variation of its area and its angular momentum\footnote{We are not going to deal with charged black holes in this paper as charge is conformally invariant and would only render our formulas longer and cumbersome. See, however, the discussion below Eq.~(\ref{MJVariation3})}. through its surface gravity and its angular velocity, respectively. For a stationary black hole of mass $M$, of angular velocity $\Phi$, of angular momentum $J$, whose surface gravity is $\kappa$, and whose surface area is $A$, the first law states that \cite{BCH},
\begin{equation}
\delta M=\frac{\kappa}{8\pi}\delta A+\Phi\delta J.
\end{equation}
The key ingredients in such a derivation are the Komar formulas giving the mass and the angular momentum of the black hole based on the timelike Killing vector $\zeta^a$ and the spacelike Killing vector $\chi^a$. These latter vectors are related to the null Killing vector $\xi^a$ through the following linear combination, $\xi^a=\zeta^a+\Phi\chi^a$ (see e.g. Ref.~\cite{Poisson} for a concise textbook presentation). Komar formulas, in turn, easily give rise to the following generalized Smarr formula \cite{Poisson}:
\begin{equation}
M-2\Phi J=\frac{\kappa A}{4\pi}.
\end{equation}
We are going to work out the explicit derivation of both the generalized Smarr formula and the first law within the conformal frame in Sec.~\ref{SecIII}.
\subsection{Second law in Jordan frame}
The second law of black hole mechanics states that the area $A$ of the event horizon never decreases. This formally reads,
\begin{equation}\label{2Law}
\delta A\geq0.
\end{equation}
Because the entropy of a black hole is identified with the area of its event horizon, the second law guarantees the satisfaction of the corresponding second law of thermodynamics. The derivation of inequality (\ref{2Law}) is based on showing that the expansion $\theta$ of the congruence of null geodesics, which measures also the variation of the horizon's surface area, satisfies $\theta\geq0$ everywhere on the horizon \cite{Hawking1971}.

\subsection{Third law in Jordan frame}
The third law of black hole mechanics states that it is impossible by any physical procedure to reduce the surface gravity $\kappa$ to zero in a finite number of steps. This is the analogue of the third law of thermodynamics which states that it is impossible to reach a zero temperature by a physical process. The derivation of this statement is based on Raychaudhuri's equation and the weak energy condition \cite{Israel}. One uses these to show that, starting from a trapped surface $S_0$, all subsequent two-sections $S(t)$ in the time $t$ are necessarily trapped. 
\section{The four laws in the conformal frame}\label{SecIII}
In this section we investigate the fate of the previous laws in the conformal frame. Given that the first two laws depend on the Killing vector, we are first going to search for the {\it true} Killing vector field $\tilde{\xi}^a$ that satisfies the Killing equation, $\tilde{\nabla}_{(a}\tilde{\xi}_{b)}=0$, in the conformal frame.

First, let us show that the new Killing vector field $\tilde{\xi}^a$ of the new frame is necessarily proportional to the Killing vector field $\xi^a$ of the original frame. In fact, start with the positive surface gravity $\kappa$ as given in the original frame by the definition (\ref{SurfaceGrav1}). Then, contract both sides of that identity by $\tilde{\xi}^b$:
\begin{align}\label{ProportionalityDerivation}
&\xi^a\tilde{\xi}^b\nabla_a\xi_b=\kappa\tilde{\xi}^b\xi_b\nonumber\\
\Longrightarrow\quad &\xi^a \Omega^{-2}\nabla_a\big(\tilde{\xi}_b\xi^b\big)-\xi^a \Omega^{-2}\xi^b\nabla_a\tilde{\xi}_b=\kappa\tilde{\xi}^b\xi_b\nonumber\\
\Longrightarrow\quad
&\xi^a \Omega^{-2}\nabla_a\big(\Omega^2\tilde{\xi}^b\xi_b\big)-\xi^a \Omega^{-2}\xi^b\tilde{\nabla}_a\tilde{\xi}_b\nonumber\\
-&\xi^a\xi^b \Omega^{-3}\big(\delta^c_a\partial_b\Omega+\delta^c_b\partial_a\Omega-g_{ab}\partial^c\Omega\big)\tilde{\xi}_c=\kappa\tilde{\xi}^b\xi_b\nonumber\\
\Longrightarrow\quad
&\xi^a\nabla_a\big(\tilde{\xi}^b\xi_b\big)=\kappa\tilde{\xi}^b\xi_b.
\end{align}
In the second step we have used $\tilde{\xi}^b\xi_b=\Omega^{-2}\tilde{\xi}_b\xi^b$, while in the third step we have introduced the covariant derivative $\tilde{\nabla}_b$ of the conformal frame, and used the conformal transformation of the Christoffel symbols, $\tilde{\Gamma}_{ab}^c=\Gamma_{ab}^c+\Omega^{-1}\left(\delta^c_a\partial_b\Omega+\delta^c_b\partial_a\Omega-g_{ab}\partial^c\Omega\right)$. In the last step we have used the fact that $\xi^a$ is a null vector, and that $\tilde{\xi}^a$ satisfies the Killing equation in the conformal frame.

Suppose now that $\tilde{\xi}^a\xi_a\neq0$. Then, according to the last line of Eq.~(\ref{ProportionalityDerivation}) we would have $\frac{\rm d}{{\rm d}\lambda}\ln(\tilde{\xi}^a\xi_a)=\kappa$, which integrates to $\tilde{\xi}^a\xi_a\sim e^{\kappa\lambda}$. Given that $\kappa$ is positive, this would make the projection of $\tilde{\xi}^a$ on the original vector $\xi^a$ rapidly diverge. Therefore, we deduce that $\tilde{\xi}^a\xi_a=0$. This, in turn, implies that there exists a scalar $f$ such that $\tilde{\xi}^a=f\xi^a$. {\large$\Box$} 

Let us now find a constraint on the scalar $f$, or at least a condition for the existence of the scalar $f$. We will show that for a Killing vector to exist in the conformal frame, when one already exists in the original frame, the conformal factor $\Omega$ needs to be invariant along the direction of the original Killing vector field. 

Start again from the fact that $\tilde{\xi}^a$ satisfies the Killing equation in the conformal frame, and use the transformation of the Christoffel symbols as given above to trade the covariant derivative in the conformal frame for the covariant derivative in the old frame,
\begin{align}\label{fConstraint}
&\nabla_a\tilde{\xi}_b+\nabla_b\tilde{\xi}_a=\frac{2}{\Omega}\left(\delta^c_a\partial_b\Omega+\delta^c_b\partial_a\Omega-g_{ab}\partial^c\Omega\right)\tilde{\xi}_c\nonumber\\
\Longrightarrow\quad&
\nabla_a\left(f\Omega^2\xi_b\right)+\nabla_b\left(f \Omega^2\xi_a\right)\nonumber\\
&\qquad\qquad\quad\;\;\,=2f \Omega\left(\xi_a\partial_b\Omega+\xi_b\partial_a\Omega-g_{ab}\xi^c\partial_c\Omega\right)\nonumber\\
\Longrightarrow\quad&
\Omega(\xi_b\partial_a f+\xi_a\partial_b f)=-2f g_{ab}\xi^c\partial_c\Omega.
\end{align}
In the last step we have used the fact that $\xi^a$ itself satisfies the Killing equation in the original frame. 
By contracting the last identity with $g^{ab}$, we have that $\Omega\xi^a\partial_a f=-4f\xi^a\partial_a\Omega$. On the other hand, by contracting both sides of the identity by $\xi^a$, we have that $\Omega\xi^a\partial_a f=-2f\xi^a\partial_a\Omega$. This can only be true if and only if $\xi^a\partial_a\Omega=0$, and hence also, $\xi^a\partial_a f=0$. {\large$\Box$}

This result means that for a conformal factor $\Omega$ which is not invariant along the direction of the original Killing vector field, no Killing vector field would exist in the conformal frame. It is not hard to show now that whatever one chooses for the multiplicative factor $f$ such that $\xi^a\partial_a f=0$, the surface gravity $\tilde{\kappa}$ in the conformal frame, given by $\tilde{\xi}^a\tilde{\nabla}_a\tilde{\xi}^b=\tilde{\kappa}\tilde{\xi}^b$, is related to the original surface gravity $\kappa$ by $\tilde{\kappa}=f\kappa$. Furthermore, one can easily check that this is true for whichever of the definitions (\ref{SurfaceGrav1}), (\ref{SurfaceGrav2}), or (\ref{SurfaceGrav3}) of the surface gravity one chooses to use.

With these preliminary results at hand, let us now check the status of the four laws of black hole mechanics in the conformal frame.

\subsection{Zeroth law in Einstein frame}
As indicated in the Sec.~\ref{SecII}, the key point in deriving the uniformity of the surface gravity over the entire horizon is to show that $\tilde{\xi}_{[a}\tilde{\nabla}_{b]}\tilde{\kappa}=0$. Actually, it is easy to convince oneself that the steps leading to this identity, as outlined below Eq.~(\ref{0Law}), all hold in the conformal frame. That this is true follows immediately from the fact that no use of Einstein equations -- which are solely responsible for spoiling conformal invariance -- has been made in such a derivation. All one uses are indeed the same equations of the original frame, only decorated everywhere by tildes.

\subsection{First law in Einstein frame}
To investigate the first law in the conformal frame, we will follow very closely the derivation given in Ref.~\cite{Poisson}. The starting point, as indicated in Sec.~\ref{SecII}, are the Komar integral formulas, written in the conformal frame, that give the mass $\tilde{M}$ and the angular momentum $\tilde{J}$ of the black hole using the timelike and spacelike Killing vectors, $\tilde{\zeta}^a$ and $\tilde{\chi}^a$, respectively \cite{Poisson},
\begin{equation}\label{KomarFormulas}
\tilde{M}=-\frac{1}{8\pi}\oint_\mathscr{H}\tilde{\nabla}^a\tilde{\zeta}^b{\rm d}\tilde{S}_{ab},\qquad \tilde{J}=\frac{1}{16\pi}\oint_\mathscr{H}\tilde{\nabla}^a\tilde{\chi}^b{\rm d}\tilde{S}_{ab}.
\end{equation}
Here, ${\rm d}\tilde{S}_{ab}=\tilde{\xi}_{[a}\tilde{N}_{b]}{\rm d}\tilde{S}$ is the oriented surface element on the horizon, while $\tilde{N}^a$ is the auxiliary null vector field such that $\tilde{\xi}^a\tilde{N}_a=-1$. The integrals (\ref{KomarFormulas}) have to be evaluated over the entire horizon $\mathscr{H}$. Now, let us go through the usual steps \cite{Poisson} as follows:
\begin{align}\label{ConfKomar}
\tilde{M}-2\tilde{\Phi}\tilde{J}&=-\frac{1}{8\pi}\oint_\mathscr{H}\tilde{\nabla}^a\left(\tilde{\zeta}^b+\tilde{\Phi}\tilde{\chi}^b\right){\rm d}\tilde{S}_{ab}\nonumber\\
&=-\frac{1}{8\pi}\oint_\mathscr{H}\tilde{\nabla}^a\tilde{\xi}^b{\rm d}\tilde{S}_{ab}\nonumber\\
&=-\frac{1}{4\pi}\oint_\mathscr{H}\tilde{N}_b\tilde{\xi}^a\tilde{\nabla}_b\tilde{\xi}^b{\rm d}\tilde{S}=\frac{\tilde{\kappa}\tilde{A}}{4\pi}.
\end{align}
Notice the important fact that in the very last step use has been made of the specific definition (\ref{SurfaceGrav1}) of the surface gravity in the conformal frame, and not the other two definitions (\ref{SurfaceGrav2}) or (\ref{SurfaceGrav3}).

As shown below Eq.~(\ref{fConstraint}), the new Killing vector field $\tilde{\xi}^a$, if it exists at all (i.e., if $\xi^a\partial_a\Omega=0$), is necessarily related to the old Killing vector field by $\tilde{\xi}^a=f\xi^a$, from which it follows that $\tilde{\kappa}=f\kappa$. Now, given that $\tilde{A}=\Omega^2 A$, the natural choice for $f$ is $f=\Omega^{-1}$. In fact, only with such a choice do we recover from Eq.~(\ref{ConfKomar}) the usual conformal transformation of a {\it geometric} mass,
\begin{equation}\label{GeometricMass}
\tilde{M}-2\tilde{\Phi}\tilde{J}=\frac{\tilde{\kappa}\tilde{A}}{4\pi}=\Omega\frac{\kappa A}{4\pi}=\Omega(M-2\Phi J).
\end{equation}
This specific conformal transformation (with a factor of $\Omega$ instead of $\Omega^{-1}$ as usual masses do) is indeed just signaling the purely geometric nature, rather than the purely material nature, of the quantity as shown in Refs.~\cite{Hammad1,Hammad2}. Given the purely geometric entities contained in the left-hand side of Komar formulas (\ref{KomarFormulas}), though, this is hardly surprising. This will not be the case, however, for the variation of such mass and angular momentum as such variation is found based on the purely material stuff that fell into the black hole.   

In fact, to find the variation of the mass and angular momentum, introduce an infinitesimal perturbation caused by an infinitesimal flow of energy-momentum carried by the tensor $\tilde{T}_{ab}$ across the horizon \cite{Poisson}. The infinitesimal variation in the conformal frame of the mass of the black hole and its angular momentum are then given by \cite{Poisson},
\begin{equation}\label{MJVariation}
\delta\tilde{M}=-\int_\mathcal{H}\tilde{T}^a_{\;\;b}\tilde{\zeta}^b{\rm d}\tilde{\Sigma}_a,\qquad \delta\tilde{J}=\int_\mathcal{H}\tilde{T}^a_{\;\;b}\tilde{\chi}^b{\rm d}\tilde{\Sigma}_a.
\end{equation}
Here, ${\rm d}\tilde{\Sigma}_a=\tilde{\xi}_a{\rm d}\tilde{S}{\rm d}\tilde{\lambda}$ is the surface element along the hypersurface of the horizon $\mathcal{H}$, while $\tilde{\lambda}$ is the non-affine parameter along the Killing vector $\tilde{\xi}^a$ before perturbation. Combining the two identities (\ref{MJVariation}), we get
\begin{equation}\label{MJVariation2}
\delta\tilde{M}-\tilde{\Phi}\delta\tilde{J}=\int{\rm d}\tilde{\lambda}\oint_\mathscr{H}\tilde{T}_{ab}\tilde{\xi}^a\tilde{\xi}^b{\rm d}\tilde{S}.
\end{equation}
At this point, we need to use Raychaudhuri's equation in the conformal frame. It is not hard to show that for a null vector $\tilde{\xi}^a$ in the conformal frame such that $\tilde{\nabla}^a\tilde{\xi}_a\tilde{\xi}^b=\tilde{\kappa}\tilde{\xi}^b$, the expansion $\tilde{\theta}$ of the geodesics congruence obeys the same Raychaudhuri equation \cite{Poisson} of the original frame:
\begin{equation}\label{Raychaudhuri}
\frac{{\rm d}\tilde{\theta}}{{\rm d}\tilde{\lambda}}=-\frac{1}{2}\tilde{\theta}^2+\tilde{\kappa}\tilde{\theta}-\tilde{\sigma}_{ab}\tilde{\sigma}^{ab}+\tilde{\omega}_{ab}\tilde{\omega}^{ab}-\tilde{R}_{ab}\tilde{\xi}^a\tilde{\xi}^b.
\end{equation}
In fact, to get this equation, one either ($i$) finds the conformal transformations of the congruence's expansion $\theta$, the shear tensor $\sigma_{ab}$ and the twist tensor $\omega_{ab}$, and then uses these conformal expressions to derive the Raychaudhuri equation as is usually done, or ($ii$) one can just pick up Raychaudhuri's equation in the original frame and decorate it by tildes everywhere given that the equation is purely geometric and does not involve Einstein's field equations.

Next, notice that both $\tilde{\theta}^2$ and $\tilde{\sigma}_{ab}\tilde{\sigma}^{ab}$ are second order in the perturbation caused by the $\tilde{T}_{ab}$, while $\tilde{\omega}_{ab}$ is zero by Frobenius' theorem because $\tilde{\xi}^a$ is null, and hence hypersurface orthogonal. Therefore, only the second and last terms on the right-hand side will be kept in Eq.~(\ref{Raychaudhuri}). Finally, one has now to use Einstein field equations to introduce $\tilde{T}_{ab}$ inside Eq.~(\ref{Raychaudhuri}). As already mentioned above, Einstein equations are not invariant under conformal transformations. In fact, in the conformal frame they read
\begin{equation}
\tilde{R}_{ab}-\frac{\tilde{R}}{2}\tilde{g}_{ab}=8\pi\Omega^2\tilde{T}_{ab}-\frac{2\tilde{\nabla}_a\!\tilde{\nabla}_b\Omega}{\Omega^2}+\tilde{g}_{ab}\!\!\left[\frac{2\tilde{\Box}\Omega}{\Omega}\!-\!\frac{3(\partial\Omega)^2}{\Omega^2}\right].
\end{equation}
Contracting both sides of these equations by $\tilde{\xi}^a\tilde{\xi}^b$, keeping in mind that $\tilde{\xi}^a$ is a null vector, and then substituting in the right-hand side of Eq.~(\ref{Raychaudhuri}), turns the latter into,
\begin{equation}\label{RaychaudhuriWithT}
\frac{{\rm d}\tilde{\theta}}{{\rm d}\tilde{\lambda}}=\tilde{\kappa}\tilde{\theta}-8\pi\Omega^2\tilde{T}_{ab}\tilde{\xi}^a\tilde{\xi}^b+\frac{2\tilde{\xi}^{a}\tilde{\xi}^b\tilde{\nabla}_a\!\tilde{\nabla}_b\Omega}{\Omega^2}.
\end{equation}
Extracting $\tilde{T}_{ab}\tilde{\xi}^{a}\tilde{\xi}^b$ from this equation and substituting it inside the right-hand side of Eq.~(\ref{MJVariation2}), we find after integration,
\begin{align}\label{ConfFirstLaw}
\delta\tilde{M}-\tilde{\Phi}\delta\tilde{J}
&=\frac{1}{8\pi\Omega^2}\int {\rm d}\tilde{\lambda}\oint_\mathscr{H}\left(\tilde{\kappa}\tilde{\theta}-\frac{{\rm d}\tilde{\theta}}{{\rm d}\tilde{\lambda}}\right){\rm d}\tilde{S}\nonumber\\
&\quad+\frac{1}{4\pi}\int {\rm d}\tilde{\lambda}\oint_\mathscr{H}\frac{\tilde{\xi}^a\tilde{\xi}^b\tilde{\nabla}_a\tilde{\nabla}_b\Omega}{\Omega^4}{\rm d}\tilde{S}\nonumber\\
&=\frac{\tilde{\kappa}}{8\pi\Omega^2}\delta\tilde{A}.
\end{align}
The last line comes only from the first double-integral on the right-hand side after performing an integration by parts and using the fact that the black hole is stationary both before and after the perturbation \cite{Poisson}. Note that we have taken out of the integrals the factor $\Omega$ because $\tilde{\xi}^a\partial_a\Omega=0$ and, as we show it in Appendix \ref{ConstantOmega} below, we have $\tilde{\xi}_{[c}\tilde{\nabla}_{d]}\Omega=0$ and therefore $\Omega$ is uniform all over the horizon. The last double-integral on the right-hand side vanishes after integrating by parts and using the fact that $\tilde{\xi}^a\partial_a\Omega=0$.

The result (\ref{ConfFirstLaw}) represents the first law of black hole mechanics in the conformal frame. Notice that the form of the first law is thus non-invariant under Weyl transformations. Indeed, if it were, we would have found instead an expression of the form $\delta\tilde{M}-\tilde{\Phi}\delta\tilde{J}=\tilde{\kappa}\delta\tilde{A}/8\pi$. It follows from this observation that if one is still willing to keep the analogy with black hole mechanics and thermodynamics, then one should identify the black hole temperature not with $\tilde{\kappa}$, but with $\tilde{\kappa}/\Omega^2$. But then, as we have seen above concerning the behavior of surface gravity when choosing $f=1/\Omega$ --- and as we will see below --- the expected behavior of temperature, which should be that of an energy, would be spoiled.

In fact, another way of finding how surface gravity should transform is through the effect on geometry of the infalling matter towards the black hole. Indeed, going back to Eq.~(\ref{MJVariation2}) and substituting there the relations $\tilde{\xi}^a=\Omega^{-1}\xi^a$, $\tilde{T}_{ab}=\Omega^{-2} T_{ab}$, $\tilde{\lambda}=\Omega\lambda$, and ${\rm d}\tilde{S}=\Omega^2{\rm d}S$, we get
\begin{align}\label{MJVariation3}
\delta\tilde{M}-\tilde{\Phi}\delta\tilde{J}&=\Omega^{-1}\int{\rm d}\lambda\oint_\mathscr{H}T_{ab}\xi^a\xi^b{\rm d}S\nonumber\\
&=\Omega^{-1}\left(\delta M-\Phi\delta J\right)\nonumber\\
&=\frac{\kappa\delta A}{8\pi\Omega}.
\end{align}
To take $\Omega$ out of the integral in the first line we have used again the uniformity of $\Omega$ over the entire horizon thanks to $\xi_{[a}\nabla_{b]}\Omega=0$. Now comparing the ratio in the last line of the result (\ref{MJVariation3}) with that of the last line in Eq.~(\ref{ConfFirstLaw}), we immediately conclude that the surface gravity transforms indeed as $\tilde{\kappa}=\kappa/\Omega$. 

Two very important remarks are here in order. The first, is that none of the equations (\ref{ConfFirstLaw}) and (\ref{MJVariation3}) shows conformal invariance. As already mentioned in the footnote of Sec.~\ref{sec:IIB}, however, the black hole's charge and its electric potential do remain invariant under the conformal transformation. This observation means that adding charge to the system just makes things worse regarding the analogy between black hole mechanics and black thermodynamics in the conformal frame. This is because no coherent picture could then be drawn from the different scaling of the various terms making the two formulas. The deep reason behind this extra issue lies in the fact that, while entropy and temperature have been translated into geometry on the right-hand side of the first law of black hole mechanics, charge and its electric potential remained in their purely ``material'' nature.

The second remark is that the conformal transformation of the variation $\delta M-\Phi\delta J$, as is displayed in the second line of Eq.~(\ref{MJVariation3}), comes with the factor $\Omega^{-1}$, whereas the conformal transformation of $M-2\Phi J$ comes with the factor $\Omega$ as shown in Eq.~(\ref{GeometricMass}). This fact stems from the two {\it different} natures of the quantities. The former has been obtained from the infalling matter through the use of the energy-momentum tensor $T_{ab}$, whereas the latter has been obtained through the geometry of the horizon. The first is thus a purely material quantity whereas the second is a purely geometric entity. This point adds additional weight to what has been stressed in Refs.~\cite{Hammad1, Hammad2}.  


\subsection{Second law in Einstein frame}
As recalled in Sec.~\ref{SecII}, the second law of black hole mechanics states that the area of the event horizon never decreases, which in turn is based on the fact that $\theta\geq0$. To find the analogue of this in the conformal frame, one can simply use the fact that $\tilde{A}=\Omega^2 A$, which, because the second law is satisfied in the original frame, immediately yields the second law in the conformal frame,
\begin{equation}\label{Conf2Law}
\delta\tilde{A}\geq0.
\end{equation}
More rigorously, however, one can use the fact that the expansion $\tilde{\theta}$ in the conformal frame is related to the expansion $\theta$ in the original frame by $\tilde{\theta}=\Omega^{-1}(\theta+2\xi^a\partial_a\ln\Omega)$ \cite{Hammad3}. Provided then only that $\theta\geq0$ in the original frame, one is guaranteed that $\tilde{\theta}\geq0$ in the conformal frame as well when the conformal factor is chosen such that $\xi^a\partial_a\Omega=0$, i.e., that a true Killing vector exists in the conformal frame.

\subsection{Third law in Einstein frame}
Given that the derivation of the third law of black hole mechanics is based on Raychaudhuri's equation {\it and} the weak energy condition, we would not be able to do justice to it in this paper without going into a detailed study of the conformal transformation of the energy conditions, which lies beyond the scope of the present paper. Therefore, we shall not attempt to derive the third law in the conformal frame here but just point out that it is not obvious at all that such a law would still hold without some constraints in the conformal frame. A rigorous study devoted to the conformal transformation of the third law will be attempted elsewhere. 
\section{The ``conformal Killing horizon"}\label{SecIV}
After having examined what the laws of black hole mechanics become under a Weyl transformation and deduced the way surface gravity should transform, the next natural task would be to confront these results with the conformal invariance usually obtained in the literature based on more ``sophisticated" methods for computing the Hawking temperature. However, given that our approach goes back to the classical origins of black hole thermodynamics, and is not concerned with any quantum considerations yet, we are going to focus in this paper only on one similar investigation. This is the work done in Ref.~\cite{JacobsonKang}. 

According to the authors of that paper, one can use the so-called ``conformal Killing vector" from which one can extract a conformally invariant surface gravity that one would identify with the temperature of the black hole. First of all, it must be noted in this regard that among the three equivalent definitions (\ref{SurfaceGrav1}), (\ref{SurfaceGrav2}), and (\ref{SurfaceGrav3}) of surface gravity, only the second answers such a requirement of conformal invariance \cite{JacobsonKang}. 

Thus, instead of looking for a true Killing vector field in the conformal frame, the authors of Ref.~\cite{JacobsonKang} proposed to use this specific ``conformal Killing vector". The latter is just the old Killing vector $\xi^a$ of the original frame, whose Killing's equation is, consequently, spoiled in the conformal frame \cite{JacobsonKang},
\begin{equation}\label{ConfKillingEq}
\tilde{\nabla}_{(a}\xi_{b)}=\tilde{g}_{ab}(\pounds_\xi\ln\Omega).
\end{equation}
The surface gravity that is identified with temperature in the conformal frame is thus the one extracted from this vector, but built using exclusively the definition (\ref{SurfaceGrav2}). In this section, however, we argue that the mere conformal invariance of such a quantity does not justify why one should identify it with the temperature of the black hole horizon. 

In fact, we find that such a surface gravity violates the zeroth law of black hole mechanics, and hence would also violate the zeroth law of thermodynamics in case one identifies it with the temperature of the black hole. In order to come to such a conclusion, one should show that over the entire horizon, the quantity, $\xi_{[a}\tilde{\nabla}_{b]}\kappa$, does not vanish as it should for a uniform surface gravity. The derivation of this claim is lengthy, so the explicit calculations are deferred to Appendix A. Therefore, identifying the surface gravity extracted from the conformal Killing vector with the black hole temperature in the conformal frame is problematic.
\section{Thought experiments in the conformal frame}\label{SecV}
Given these unexpected results from the way black hole mechanics transforms under Weyl mapping, one might wonder what the usual heuristic and intuitive thought experiments that give rise to the familiar formulas of black hole thermodynamics would imply in the conformal frame. We are going to examine a thought experiment that leads to the usual entropy area law first, and then investigate a thought experiment that leads to Hawking temperature.

\subsection{Entropy}
Imagine dropping a small cubic box full of hot gas of temperature $T$ into a black hole of mass $M$. Suppose the box of gas has a linear size $L$ and a mass $m$. The initial state then consists of the box of gas and the black hole of radius $R=2M$, while the final state consists of the single, but larger, black hole. The box of gas will merge with the black hole when its proper distance $\rho$ from the horizon is of order $L$. On the other hand, the gas inside the box is in thermal equilibrium at temperature $T$, so we assume the box is at least as large as the thermal wavelength of the gas it contains; that is, $L\sim1/T$. Therefore, once the box disappears behind the horizon, there will be an entropy loss of, $\Delta\mathcal{S}_{\rm gas}\sim-m/T\sim-mL$. 

Now let us see what happens from the point of view of the conformal frame. Under the Weyl transformation (\ref{Weyl}), the mass of the gas inside the box becomes $\tilde{m}=m/\Omega$, while the linear size of the box becomes $\tilde{L}=\Omega L$. Thus, the resulting loss of entropy in the conformal frame should be,
\begin{equation}\label{GasEntropyLoss}
\Delta\tilde{\mathcal{S}}_{\rm gas}=-\tilde{m}\tilde{L}=-mL=\Delta\mathcal{S}_{\rm gas}.
\end{equation}
According to this, the entropy of the gas should thus be conformally invariant. This is understandable given that entropy is a measure of the number of degrees of freedom of matter and conformal transformations should not in principle alter such a number of degrees of freedom.

Let us now check things on the side of the black hole. For simplicity, we consider a Schwarzschild black hole of mass $M$ whose metric ${\rm d}\tilde{s}$ in the conformal frame reads, ${\rm d}\tilde{s}=-\Omega^2(1-\frac{2M}{r}){\rm d}t^2+\Omega^2(1-\frac{2M}{r})^{-1}{\rm d}r^2+\Omega^2r^2{\rm d}\vartheta^2$. Here, $d\vartheta^2$ is the metric on the unit sphere. The proper distance $\tilde{\rho}$ of the box from the Schwarzschild horizon in the conformal frame is thus,
\begin{equation}\label{ProperDistance}
\tilde{\rho}=\int_{2M}^{2M+\delta r}\frac{\Omega{\rm d}r}{\sqrt{1-2M/r}}\sim \Omega\sqrt{M\delta r}.
\end{equation}
The box of gas will coalesce with the black hole in the conformal frame when its proper distance $\tilde{\rho}\sim\tilde{L}$. According to Eq.~(\ref{ProperDistance}), this would happen then for $\delta r\sim L^2/M$. When the box of mass $m$ reaches the coordinate distance $r=2M+\delta r$, it will thus be considered as being part of the black hole. The latter will then appear, for a distant observer at infinity, as having gained a redshifted mass of,
\begin{equation}\label{MassGain}
\Delta\tilde{M}\sim \tilde{m}\Omega\sqrt{1-\frac{2M}{2M+\delta r}}\sim\frac{\Omega\tilde{m}L}{M}=\frac{mL}{M}.
\end{equation}
The change in the horizon area after this absorption of the box by the black hole in the conformal frame would therefore have to be, 
\begin{equation}\label{AreaIncrease}
\Delta\tilde{A}\sim M\Delta\tilde{M}\sim mL\sim\Delta A,
\end{equation}
which is not at all the way surface areas transform under conformal transformations. Furthermore, to preserve the second law of thermodynamics, we must have an increase $\Delta\tilde{\mathcal{S}}_{\rm BH}$ of the black hole entropy such that, $\Delta\tilde{\mathcal{S}}_{\rm BH}\sim\Delta \tilde{\mathcal{S}}_{\rm gas}\sim \Delta\mathcal{S}_{\rm gas}\sim\Delta \mathcal{S}_{\rm BH}$. If entropy should really be proportional to area in the conformal frame in accordance with the area law of the original frame, and area does not remain invariant after a Weyl transformation, how come that entropy remains invariant like it does here after such a transformation? 

Actually, just like what we found for the entropy of the gas inside the box, the black hole entropy should intuitively remain conformally invariant if it really represents the number of degrees of freedom of the black hole. The fact that we just confirmed this expected behavior goes back to the non-geometric character of our intuitive argument followed here, in complete contrast to what the purely geometric approach of Sec.~\ref{SecIII} implies.

It turns out that this apparent contradiction between the two approaches is not restricted to entropy. As we will see below, intuitive arguments based solely on the material part leads to the same discordance between geometry-based and matter-based approaches.

\subsection{Temperature}
For the problem of temperature, let us follow the heuristic approach suggested by Hawking, that consists in examining vacua just outside the horizon. For that purpose let us view the vacuum as a sea of virtual pairs of particles and antiparticles of energy $|E|$, in absolute value, momentarily at rest at a coordinate distance $\delta r$ from the horizon. The proper time for the negative energy particle to reach the horizon in the conformal frame, i.e., the lifetime of the pair is, as in Eq.~(\ref{ProperDistance}), $\tilde{\tau}\sim\Omega\sqrt{M\delta r}$. Hence, the energy of the pair is roughly the inverse of this lifetime,
\begin{equation}\label{PairEnergy}
|\tilde{E}|\sim\frac{1}{\Omega\sqrt{M\delta r}}.
\end{equation}
The energy of the pair, and hence the temperature, being at the coordinate distance $2M+\delta r$, will be redshifted for an observer at infinity and takes the value, 
\begin{equation}\label{PairEnergy}
k_B\tilde{T}\sim|\tilde{E}|_\infty\sim\frac{1}{\Omega\sqrt{M\delta r}}\Omega\sqrt{1-\frac{2M}{2M+\delta r}}\sim\frac{1}{M}\sim k_BT.
\end{equation}
Here, $k_B$ is the Boltzmann constant. The Hawking temperature in the conformal frame, according to this result, should then be the same as that of the original frame. 

This is understandable as being due to the fact that the usual $\Omega^{-1}$ rescaling of energy and the usual $\Omega$ rescaling of time causing the redshift cancel each other. Another way to see why this still makes sense is by observing that a conformal transformation does not affect the central {\it material} mass $M$ of the black hole. Therefore, given that $T\sim1/M$, temperature, as for entropy, should not be affected either. However, from the point of view of an observer -- or a detector -- temperature is, up to the Boltzmann constant, equivalent to energy and, as such, should scale like $\Omega^{-1}$ as already found above by considering the surface gravity.

Having again found two different results does not mean that one result should be wrong and the other correct. Both results are actually correct. One only needs to specify from which point of view one is considering black hole thermodynamics: is it from the geometric viewpoint or from the material viewpoint? If it is the former then one would better call it ``geometric thermodynamics" whereas if it is the latter then ``material thermodynamics" would be more appropriate. Only in ``material thermodynamics" are the equations invariant. If a conclusion is to be drawn from these results, it would be that one should be careful when using black hole thermodynamics to deepen our knowledge about spacetime.

\section{Summary \& discussion}\label{Conclusion}
The conformal behavior of black hole thermodynamics has been investigated based on the conformal behavior of the laws of black hole mechanics. We have dealt solely with stationary black holes as they are the most important for a conceptual understanding of black hole thermodynamics and for which an easier meaningful and useful concept of temperature can be defined. The conformal invariance of Hawking temperature found in some of the literature on the subject does not hold according to this approach. The latter shows indeed that the ``special" surface gravity that supports such a result cannot be identified with temperature in the first place. 

The study we did in this paper implies that any investigation of the conformal behavior of a physical phenomenon involving spacetime and matter is very subtle. Recall that, historically, black hole thermodynamics was born from the classical considerations of black hole mechanics. The ``sophisticated'' methods that recover Hawking temperature and add support to this analogy came later. Therefore, the fact that a conformally invariant temperature based on the ``sophisticated" methods in the literature, devised for recovering black hole thermodynamics, is also in conflict with what is found here, might only suggest that a key insight is still missing. We shall come back to this point shortly below.

Now, finding that Hawking temperature is invariant under a Weyl transformation should not actually leave one indifferent. In fact, if the black hole temperature remains invariant under space and time rescaling, one cannot merely invoke Dicke's argument \cite{Dicke} for the equivalence of the two frames. Recall that Dicke's argument consists in taking into account the effect on clocks and rulers of the conformal transformation (\ref{Weyl}) of the metric. The proper time $\tau$ measured by a given clock in one frame would be measured as being $\tilde{\tau}=\Omega\tau$  by the same clock in the new frame. Similarly, the proper distance $\ell$ measured by a ruler in the original frame would be measured as being $\tilde{\ell}=\Omega\ell$ by the same ruler in the new frame. This is because both the clock and the ruler are objects whose physical manifestation is conditioned by the specific spacetime arena they are immersed in. Now given that all our measurements are, in some sense, reducible to measurements of time and space (and mass), this similarity in the rescaling of space and time, i.e., of our units of measurements, would {\it a priori} make all our measurements indifferent towards the conformal transformations. Take energy (or temperature measured by a thermometer) as an example of a physical quantity. As energy has dimensions of inverse time, it is clear that energy should scale with a factor of $\Omega^{-1}$ under a Weyl transformation. On the other hand, in terms of clocks, energy could be thought of, up to Planck's constant, as the frequency, i.e., inverse period, of the photon released by an electron jumping from one energy level to another. As the clock's time itself is stretched, though, the factor of $\Omega^{-1}$ appearing in the new energy would simply not be perceived by the observer, for the unit of frequency, the hertz, itself has been affected and changed into $\widetilde{\rm hertz}=\Omega^{-1}$hertz, so to speak. This argument is valid for any other physical quantity that gets affected by the conformal transformation. Thus, it is easy to see that a physical quantity, like speed, which is dimensionless in  natural units, is indeed not affected by a conformal transformation, for the simultaneous rescalings of space and time just cancel each other in the ratio. Another example, of higher importance, which is dimensionless and thus could safely be invariant without clashing with Dicke's argument, is entropy. However, if a physical quantity does have the dimensions of space, time, or energy, and yet remains invariant under a conformal transformation, any observer in the conformal frame would perceive differently the same physical quantity. This stems from the fact that while the measuring devices -- clocks and rulers -- are affected by the transformation, nothing in the physical quantity itself changes accordingly to compensate for the change in those measuring devices. A good example here would be the Hawking temperature. 

In summary then, Dicke's argument would only work whenever a physical quantity is not invariant under a Weyl transformation --- provided that such a quantity transforms like space, time, or energy. For then a rescaling of space and time would not be detected by the observer's rescaled clocks and rulers. The conflict would arise instead for a physical quantity --- like energy --- that appears conformally invariant. However, the aim of our present paper was not to provide a definitive conclusion to the debate about whether a conformal transformation of spacetime leaves the physics invariant; a debate that has not yet been settled \cite{ConformalIssue}. To do justice to such a debate indeed it is not sufficient to rely on a single specific example --- here black hole thermodynamics. An in-depth investigation, in which the root of the issue, which is the interaction between spacetime and matter, is expected to give rise to novel insights and a general and definitive conclusion about the issue. Such an investigation will be attempted elsewhere. Suffice it to point out here that a non-invariance of Hawking temperature under conformal transformations is not at all in conflict with Dicke's argument, but, more importantly, it is called for in order for the equivalence of the two frames to hold.

Finally, coming back to the main theme of the paper, it is clearly not at all surprising that the original analogy between black hole mechanics and thermodynamics does not hold anymore in the conformal frame. By performing a conformal transformation one indeed does not turn a mass that created the solution (the deformation of spacetime) into a conformally transformed mass that creates a conformally transformed solution with everything around also conformally transformed. Instead, everything around is conformally transformed, including geometry, but the central mass of the black hole remains intact as nothing is known about it once it becomes hidden behind the horizon. Furthermore, because matter and geometry behave differently under a Weyl transformation, it is no wonder why the same physical phenomenon appears to behave differently whether one focuses on the matter part or on the geometric part. 

If black hole thermodynamics is viewed purely from the geometric side, i.e., through the lens of black hole mechanics, then this phenomenon can never be conformally invariant. If, on the other hand, this phenomenon is viewed from the material side, which has historically helped turn the analogy between the mechanics and the thermodynamics into an identity, then yes it is conformally invariant. Clashes only appear when one mixes the two fundamental concepts. This observation is further supported by the discussion we had below Eq.~(\ref{MJVariation3}) about the effect of including charge into the first law. This result is, as explained in the Introduction, very reminiscent and deeply linked to what has already been found in Refs.~\cite{Hammad1,Hammad2} concerning the fundamental concept of quasilocal energy. The question that imposes itself here is then: To what extent should one rely on this analogy between black hole mechanics and thermodynamics when trying to advance our understanding of spacetime if a mere conformal transformation is able to destroy such an analogy? The simple answer is that we do not know. The results presented here merely help bring into light this fundamental issue but cannot provide an answer to such question. At best, they just suggest that one should be careful whenever attempting to build anything fundamental about spacetime based on this phenomenon.

Our investigation here suggests that the heart of the problem goes to a single key point, which is that Einstein equations are displaying a clear dichotomy between matter and geometry. These two are related only through Newton's constant. As such, the effect of a conformal transformation is then to unravel the true nature of a given quantity in any physical phenomenon involving spacetime and matter, including, as we saw, black hole thermodynamics.

\begin{acknowledgments}
F.H. gratefully acknowledges the support of the Natural Sciences \& Engineering Research Council of Canada (NSERC) (Grant No.~2017-05388) as well as the STAR Research Cluster of  Bishop's University. \'E.M. is partly supported by NSERC's Undergraduate Student Research Awards (USRA No.~2018-523478). P.L. gratefully acknowledges support from the Fonds de Recherche du Qu\'ebec - Nature et Technologies (FRQNT) via a Grant from the Programme de Recherches pour les Chercheurs de Coll\`ege and from  the STAR Research Cluster of  Bishop's University. The authors are grateful to the anonymous referee for his/her valuable comments that helped improve the clarity of the manuscript.
\end{acknowledgments}

\appendix

\section{$\xi_{[\rho}\tilde{\nabla}_{\sigma]}\kappa\neq0$}\label{NonUniformKappa}
\renewcommand{\theequation}{\thesection.\arabic{equation}}
In this appendix we are going to prove that the surface gravity based on the ``conformal Killing vector" cannot be uniform over the conformal Killing horizon.
In the course of this derivation, various identities are going to be useful, so we start by writing these down. 

The first useful identity is the one we easily deduce by using Eq.~(\ref{ConfKillingEq}) and Frobenius' theorem in the conformal frame that states that, $\xi_{[a}\tilde{\nabla}_b\xi_{c]}=0$,
for any hypersurface orthogonal vector $\xi^a$, which is specifically the case for the ``conformal Killing vector" of Eq.~(\ref{ConfKillingEq}). Then, expanding this identity, and then using Eq.~(\ref{ConfKillingEq}), we easily arrive at,
\begin{align}\label{EqA}
\xi_{[a}\tilde{\nabla}_{b]}\xi_{c}=-&\frac{1}{2}\xi_{c}\tilde{\nabla}_{a}\xi_{b}\nonumber\\
+&\frac{1}{2}(\xi_a \tilde{g}_{bc}+\xi_c\tilde{g}_{ab}-\xi_b\tilde{g}_{ca})(\pounds_\xi\ln\Omega).
\end{align}
The second useful identity one extracts from Eq.~(\ref{ConfKillingEq}), after using the fact that, $2\tilde{\nabla}_{[a}\tilde{\nabla}_{b]}\xi_c=\tilde{R}_{abcd}\xi^d$, as well as the identity, $\tilde{R}_{[abc]d}=0$, satisfied by the Riemann tensor, is,
\begin{align}\label{EqB}
\tilde{\nabla}_{a}\tilde{\nabla}_{b}\xi_{c}&=\tilde{R}_{cbad}\xi^d\nonumber\\
&+(\tilde{g}_{ac}\tilde{\nabla}_b+\tilde{g}_{bc}\tilde{\nabla}_a-\tilde{g}_{ab}\tilde{\nabla}_c)(\pounds_\xi\ln\Omega).
\end{align}
By referring to the same equations derived in Ref.~\cite{Wald}, we easily notice the extra factor that makes all formulas depart from those of a true Killing vector. This extra factor vanishes indeed for a true Killing vector thanks to the requirement, $\xi^a\partial_a\Omega=0=\pounds_\xi\ln\Omega$, as we saw above. 

The next step is to go back to the specific definition (\ref{SurfaceGrav2}), chosen for the surface gravity in Ref.~\cite{JacobsonKang}, and apply to it the operator $\xi_{[c}\tilde{\nabla}_{d]}$. After some manipulations similar to those presented in Ref.~\cite{Wald}, but based on a repeated use of Eqs.~(\ref{ConfKillingEq}), (\ref{EqA}), and (\ref{EqB}) instead, we get,
\begin{align}\label{Intermediate}
\xi_{a}\xi_{[c}\tilde{\nabla}_{d]}\kappa=\;\;&\xi^b\tilde{R}_{ab[d}^{\;\;\;\;\;\;\;e}\xi_{c]}\xi_e\nonumber\\
-&\big(\xi_{[c}\tilde{\nabla}_{|a|}\xi_{d]}+\kappa g_{a[d}\xi_{c]}\big)\left(\pounds_\xi\ln\Omega\right)\nonumber\\
-&\left(\xi_a\xi_{[c}\tilde{\nabla}_{d]}-\xi_{[c}g_{d]a}\xi^b\tilde{\nabla}_b\right)(\pounds_\xi\ln\Omega).
\end{align}

Next, applying the operator $\xi_{[d}\tilde{\nabla}_{e]}$ to Eq.~(\ref{EqA}), and then using again Eq.~(\ref{EqA}) repeatedly, leads to a lengthy expression which greatly simplifies when contracting both sides of that expression with $g^{ce}$. The final result is the following identity, 
\begin{align}\label{Intermediate2}
-\xi^e\tilde{R}_{de[b}^{\;\;\:\:\;\:\;\;c}\xi_{a]}\xi_c&=\xi_{[a}\tilde{R}_{b]}^{\;\;\;e}\xi_d\xi_e\nonumber\\
&-\big[2\xi_{[a}\tilde{\nabla}_{|d|}\xi_{b]}+\xi_{[b}\tilde{\nabla}_{a]}\xi_d+\kappa\xi_{[a}\tilde{g}_{b]d}\nonumber\\
&-2\xi_d\tilde{\nabla}_a\xi_b+2\xi_d g_{ab}(\pounds_\xi\ln\Omega)\big](\pounds_\xi\ln\Omega)\nonumber\\
&+\big(\xi_d\xi_{[a}\tilde{\nabla}_{b]}+\tilde{g}_{d[b}\xi_{a]}\xi^e\tilde{\nabla}_e\big)(\pounds_\xi\ln\Omega).
\end{align}
Finally, substituting this expression of the first term into the right-hand side of Eq.~(\ref{Intermediate}) and then contracting both sides of the resulting equation by the auxiliary null vector $N^a$, which satisfies $N^a\xi_a=-1$, yields,
\begin{align}\label{Intermediate3}
&\xi_{[c}\nabla_{d]}\kappa=-\xi_{[c}\tilde{R}_{d]}^{\;\;\;
a}\xi_a\nonumber\\
&-\left[(N^a\tilde{\nabla}_a\xi_{[d})\xi_{c]}+(\xi_{[d}\tilde{\nabla}_{c]}\xi_a)N^a+2\tilde{\nabla}_c\xi_d\right](\pounds_\xi\ln\Omega)\nonumber\\
&+\left[2\tilde{g}_{cd}(\pounds_\xi\ln\Omega)+2\xi_{[d}\tilde{\nabla}_{c]}\right](\pounds_\xi\ln\Omega).
\end{align}

At this point, one could either invoke Einstein equations and show that the very first term on the right-hand side of Eq.~(\ref{Intermediate3}) vanishes, or else relate that term to the twist one-form given by, $\tilde{\omega}_a=\tilde{\epsilon}_{abcd}\xi^b\tilde{\nabla}^c\xi^d$, and get away without invoking Einstein equations. Given that Einstein equations spoil conformal invariance, we prefer to base our proof on purely geometric arguments that are conformally invariant. Thus, we are going to use the second purely geometric method. 

First, using the identity $\tilde{\epsilon}^{abcd}\tilde{\epsilon}_{aefg}=-6\delta^{[b}_e\delta^{c}_f\delta^{d]}_g$ satisfied by the totally antisymmetric tensor in the conformal frame, we easily check that,
\begin{equation}\label{IntermediateTwist}
\tilde{\epsilon}^{abcd}\tilde{\nabla}_c\tilde{\omega}_d=6\tilde{\nabla}_c(\xi^{[c}\tilde{\nabla}^a\xi^{b]}).
\end{equation}
Now using the ``spoiled" Killing equation (\ref{ConfKillingEq}) satisfied by the ``conformal Killing vector" $\xi^a$, together with identity (\ref{EqB}), we have the following three identities,
\begin{align}\label{IntermediateTwist2}
&\xi^{[c}\tilde{\nabla}^{a}\xi^{b]}=\tfrac{1}{3}(\xi^c\tilde{\nabla}^a\xi^{b}+\xi^b\tilde{\nabla}^c\xi^{a}+\xi^a\tilde{\nabla}^b\xi^{c})\nonumber\\
&\qquad\qquad-\tfrac{1}{3}\left(\xi^c \tilde{g}^{ab}+\xi^b\tilde{g}^{ca}+\xi^a\tilde{g}^{bc}\right)(\pounds_\xi\ln\Omega),\nonumber\\[5pt]
&\tilde{\nabla}_c(\xi^c\tilde{\nabla}^a\xi^b)=\big(4\tilde{\nabla}^a\xi^b+2\xi^{[b}\tilde{\nabla}^{a]}+\tilde{g}^{ab}\xi^{c}\tilde{\nabla}_c\big)(\pounds_\xi\ln\Omega)\nonumber\\[5pt]
&\tilde{\nabla}_c(\xi^b\tilde{\nabla}^c\xi^a+\xi^a\tilde{\nabla}^b\xi^c)=-2\xi^{[b}\tilde{R}^{a]}_{\;\;\;c}\xi^c+2\tilde{\nabla}^b\xi^a(\pounds_\xi\ln\Omega)\nonumber\\
&\qquad\qquad\qquad\qquad\qquad\;\;+\big(2\xi^a\tilde{\nabla}^b-4\xi^{[b}\tilde{\nabla}^{a]}\big)(\pounds_\xi\ln\Omega).
\end{align}
Combining these identities and substituting them into the right-hand side of Eq.~(\ref{IntermediateTwist}), we deduce that,
\begin{align}\label{IntermediateTwist3}
\tilde{\nabla}_{[a}\tilde{\omega}_{b]}&=-\tfrac{1}{4}\tilde{\epsilon}_{abcd}\tilde{\epsilon}^{cdef}\tilde{\nabla}_e\tilde{\omega}_f\nonumber\\
&=-\tilde{\epsilon}_{abcd}\left[\xi^{[c}\tilde{R}^{d]}_{\;\;e}\xi^e+\big( 2\tilde{\nabla}^{c}\xi^{d}+\xi^{[c}\tilde{\nabla}^{d]}\big)\right](\pounds_\xi\ln\Omega).
\end{align}
Here, the product $\tilde{\epsilon}^{abcd}\tilde{\epsilon}_{abef}=-4\delta^{[c}_e\delta^{d]}_f$ of the antisymmetric has been used. We can now express the first term of the second line in terms of the twist one-form by using again the product $\tilde{\epsilon}^{abcd}\tilde{\epsilon}_{abef}$ of the antisymmetric tensor as follows,
\begin{align}\label{IntermediateTwist4}
\xi^{[c}\tilde{R}^{d]}_{\;\;e}\xi^e=&+\tfrac{1}{4}\tilde{\epsilon}^{cdab}\tilde{\nabla}_{[a}\tilde{\omega}_{b]}\nonumber\\
&-\big(\xi^{[c}\tilde{\nabla}^{d]}+2\tilde{\nabla}^{[c}\xi^{d]}\big)(\pounds_\xi\ln\Omega).
\end{align}
Finally, substituting the right-hand side of this identity into the right-hand side of Eq.~(\ref{Intermediate3}), and then using Eq.~(\ref{ConfKillingEq}) together with $\tilde{\nabla}_{c}\xi_{d}=\tilde{\nabla}_{(c}\xi_{d)}+\tilde{\nabla}_{[c}\xi_{d]}$, provides the final sought identity,
\begin{align}\label{FinalIdentity}
&\xi_{[c}\nabla_{d]}\kappa=-\tfrac{1}{4}\tilde{\epsilon}_{cdab}\tilde{\nabla}^{[a}\tilde{\omega}^{b]}\nonumber\\
&-\Big[(N^a\tilde{\nabla}_a\xi_{[d})\xi_{c]}+(\xi_{[d}\tilde{\nabla}_{c]}\xi_a)N^a +\xi_{[c}\tilde{\nabla}_{d]}\Big](\pounds_\xi\ln\Omega).
\end{align}
It is clear from this identity that, although the hypersurface orthogonal vector $\xi^a$ guarantees that the very first term on the right-hand side in this last result is zero, all the remaining terms on the right would not allow the left-hand side to vanish. 

The other possibility could be that the combination of those remaining terms happens to vanish as well, even when assuming $\pounds_\xi\ln\Omega\neq0$. However, if the left-hand side of Eq.~(\ref{FinalIdentity}) vanishes identically, as does the first term on the right-hand side, then the remaining terms on the right-hand side should also vanish identically. So let us put the sum of these terms equal to zero. The resulting equation is a first order differential equation in $\pounds_\xi\ln\Omega$:
\begin{multline}\label{FinalIdentity2}
\Big[(N^a\tilde{\nabla}_a\xi_{[d})\xi_{c]}+(\xi_{[d}\tilde{\nabla}_{c]}\xi_a)N^a +\xi_{[c}\tilde{\nabla}_{d]}\Big](\pounds_\xi\ln\Omega)=0.
\end{multline}
Now contract both sides of this equation with $\xi^c N^d$. This gives,
\begin{equation}\label{FinalIdentity3}
\frac{\rm d}{{\rm d}\lambda}(\pounds_\xi\ln\Omega)+(\pounds_\xi\ln\Omega)^2+2\kappa(\pounds_\xi\ln\Omega)=0.
\end{equation}
This is of the Bernoulli type differential equation, whose solution is given in terms of an arbitrary constant of integration $C$ as,
\begin{equation}\label{FinalIdentity4}
\pounds_\xi\ln\Omega=\frac{2\kappa}{e^{2\lambda\kappa+C}-1}.
\end{equation}
Solving again for $\Omega$, we find,
\begin{equation}\label{FinalIdentity5}
\Omega\sim e^{-2\kappa\lambda}\left(1-e^{2\lambda\kappa+C}\right).
\end{equation}
This shows that not only the factor $\Omega$ would have to rapidly vanish on the horizon, but that it is not even guaranteed to remain positive.
Therefore, unless $\pounds_\xi\ln\Omega=0$ the right-hand side of Eq.~(\ref{FinalIdentity}) does not vanish. This completes the proof. {\large{$\Box$}}

Notice that the terms responsible for preventing the left-hand side from vanishing, and hence from preventing the surface gravity from being uniform, are all built from the factor $\pounds_\xi\ln\Omega$ and its first derivatives. It is, as emphasized above, specifically solely this factor that prevents the vector $\xi^a$ from being a true Killing vector. In addition, as shown in Sec.~\ref{SecIII}, the vanishing of this term is a necessary condition on the conformal factor one applies to define one's conformal transformation to still be able to get a stationary spacetime at the end, and hence define a stationary black hole temperature.   
\\\\
\section{$\tilde{\xi}_{[c}\tilde{\nabla}_{d]}\Omega=0$}\label{ConstantOmega}
\renewcommand{\theequation}{\thesection.\arabic{equation}}
In this appendix we are going to prove that, provided $\Omega$ satisfies the condition $\xi^a\nabla_a\Omega=0$, which guarantees the existence of a Killing vector in the conformal frame, the factor $\Omega$ is also constant all over the horizon. 

Let us apply the operator $\tilde{\xi}_{[c}\tilde{\nabla}_{d]}$ to the condition $\tilde{\xi}^a\nabla_a\Omega=\tilde{\xi}^a\tilde{\nabla}_a\Omega=0$. We have,
\begin{align}
&\tilde{\xi}_{[c}\tilde{\nabla}_{d]}\tilde{\xi}^a\tilde{\nabla}_a\Omega=0\nonumber\\
\Longrightarrow\quad
&\tilde{\xi}^a\tilde{\xi}_{[c}\big(\tilde{\nabla}_{d]}\tilde{\nabla}_a\Omega\big)+\big(\tilde{\xi}_{[c}\tilde{\nabla}_{d]}\tilde{\xi}^a\big)\tilde{\nabla}_a\Omega=0\nonumber\\
\Longrightarrow\quad
&\tilde{\xi}^a\tilde{\xi}_{[c}\big(\tilde{\nabla}_{|a|}\tilde{\nabla}_{d]}\Omega\big)-\tfrac{1}{2}\tilde{\xi}^a\tilde{\nabla}_{c}\tilde{\xi}_{d}\tilde{\nabla}_a\Omega=0\nonumber\\
\Longrightarrow\quad
&\tilde{\xi}^a\tilde{\nabla}_a\big(\tilde{\xi}_{[c}\tilde{\nabla}_{d]}\Omega\big)-\tilde{\xi}^a\big(\tilde{\nabla}_a\tilde{\xi}_{[c}\big)\tilde{\nabla}_{d]}\Omega=0\nonumber\\
\Longrightarrow\quad
&\tilde{\xi}^a\tilde{\nabla}_a\big(\tilde{\xi}_{[c}\tilde{\nabla}_{d]}\Omega\big)-\tilde{\kappa}\tilde{\xi}_{[c}\tilde{\nabla}_{d]}\Omega=0.
\end{align}
In the third line we have used Eq.~(\ref{EqA}) together with $\tilde{\xi}^a\tilde{\nabla}_a\Omega=0$, in the fourth line we have used once more $\tilde{\xi}^a\tilde{\nabla}_a\Omega=0$ to eliminate the second term and rearranged the first by extracting a total derivative, and in the last line we have used the definition (\ref{SurfaceGrav1}) for the surface gravity $\tilde{\kappa}$.  

Suppose now that $\tilde{\xi}_{[c}\tilde{\nabla}_{d]}\Omega\neq0$. Then, according to the last line, we could integrate such differential equation to obtain,
\begin{equation}
\tilde{\xi}_{[c}\tilde{\nabla}_{d]}\Omega\sim e^{\tilde{\kappa}\tilde{\lambda}},
\end{equation}
which would rapidly diverge. Therefore, we conclude that $\tilde{\xi}_{[c}\tilde{\nabla}_{d]}\Omega=0$. This completes the proof. {\large{$\Box$}}


\end{document}